\PassOptionsToPackage{table}{xcolor}
\documentclass[sigconf]{acmart}

\usepackage{color}

\usepackage{array}
\renewcommand{\arraystretch}{1.25}
\usepackage{booktabs}
\usepackage{multirow}
\usepackage{makecell}
\usepackage[table]{xcolor}

\usepackage{colortbl}
\newcommand{\tg}{\cellcolor{gray!20}}
\newcommand{\mr}[1]{\multirow{3}{*}{#1}}% 3-row multirow (Tables 3 and 5)
\newcommand{\hv}[1]{\multicolumn{1}{c}{#1}}% centered header cell (Tables 3 and 5)
\newcommand{\y}{\scalebox{0.75}{$\checkmark$}}% checkmark scaled to digit height (Table 4)
\newcommand{\n}{}% no improvement: left blank (Table 4)
\usepackage{multirow}
\usepackage{array}
\usepackage{makecell}
\usepackage{amsmath}
\usepackage{setspace}
\usepackage[capitalise]{cleveref}
\usepackage{xcolor}
\AtBeginDocument{%
  \providecommand\BibTeX{{%
    \normalfont B\kern-0.5em{\scshape i\kern-0.25em b}\kern-0.8em\TeX}}}

\copyrightyear{2027}
\acmYear{2027}
\setcopyright{cc}
\setcctype{by}
\acmConference[ASPDAC '27]{32nd Asia and South Pacific Design Automation Conference}{January 25--28, 2027}{Tokyo, Japan}
\acmBooktitle{32nd Asia and South Pacific Design Automation Conference (ASPDAC '27), January 25--28, 2027, Tokyo, Japan}
\acmDOI{10.1145/3840404.3860925}
\acmISBN{979-8-4007-2923-2/2027/01}
  
\begin{document}

\title[LLM-Assisted Verilog Code Refinement for Efficient RTL Clock Gating]
      {LLM-Assisted Verilog Code Refinement for Efficient \\ RTL Clock Gating}
%%
%% The "author" command and its associated commands are used to define
%% the authors and their affiliations.
%% Of note is the shared affiliation of the first two authors, and the
%% "authornote" and "authornotemark" commands
%% used to denote shared contribution to the research.

\author{Geuna Chang}
\affiliation{%
  \institution{Korea Advanced Institute of Science and Technology}
  \city{Daejeon}
  \country{Republic of Korea}}
\email{chang8642@kaist.ac.kr}

\author{Seohyun Kim}
\affiliation{%
  \institution{Korea Advanced Institute of Science and Technology}
  \city{Daejeon}
  \country{Republic of Korea}}
\email{shkim1122@kaist.ac.kr}

\author{Youngsoo Shin}
\affiliation{%
  \institution{Korea Advanced Institute of Science and Technology}
  \city{Daejeon}
  \country{Republic of Korea}}
\email{youngsoo@kaist.edu}

\begin{abstract}
RTL clock gating is a popular technique for reducing clock power at the early design stage. Iterative updates of Verilog code, clock-gating inference, and analysis of clock-gating efficiency (CGE) are time-consuming and pose challenges to RTL designers. LLM-assisted automatic refinement of Verilog code for improving CGE is proposed. Since CGE and power depend on switching activity, LLMs generate Idle, Typical, and Peak workloads of the design, one of which is selected by the user as the optimization target. Four LLM clock-gating agents, each given a guideline for a particular clock-gating technique, then sequentially refine the Verilog code, focusing on the registers with low flop-gating efficiency (FGE) under the selected workload. Code updates are accepted only if they are functionally equivalent and reduce post-synthesis power without violating area and timing constraints. Experimental results with five RTL designs show an average FGE improvement of 12.1\%p and an average power reduction of 23.6\% (up to 59.3\%) under the selected workloads, compared to the original Verilog code, outperforming a commercial tool and an LLM baseline, which achieve average power reductions of only 4.1\% and 8.6\%, respectively.
\end{abstract}


\maketitle

\section{Introduction}\label{sec1}

%------------------------------------------------------
% RTL clock gating
%-------------------------------------------------------
% {\color{blue}Clock-gating is a popular method to reduce power consumption \cite{cg-power1}. It is typically specified at the RTL level, where functional dependencies among signals are most apparent. \Cref{fig1}(a) illustrates a load-enable register, a common RTL clock gating, where the gating function EN is explicitly coded; a synthesis tool replaces the resulting feedback multiplexer with an integrated clock-gating (ICG) cell. \Cref{fig1}(b) shows another example, a combinational loop, where the gating condition is extracted from the logic whereas the condition is explicit in \Cref{fig1}(a).}

%{\color{blue}RTL clock gating is a popular method to reduce power consumption \cite{cg-power1}. 
% Among the design stages at which it can be applied, RTL clock gating often offers the largest power saving among clock-gating techniques \cite{tutorial}. 
% It allows synthesis tools to infer integrated clock-gating (ICG) cells from gating conditions in the RTL code.
%\Cref{fig1}(a) illustrates a load-enable register, a frequent target of RTL clock gating. Where the gating function EN is explicitly coded, a synthesis tool first implements feedback multiplexer and then infers an ICG cell.
% to replace the feedback multiplexer. 
%\Cref{fig1}(b) shows another example, a combinational loop, where the gating condition is extracted from the logic whereas the condition is explicit in (a).}

\begin{figure}[t]
    \centering
    \includegraphics[scale=0.8]{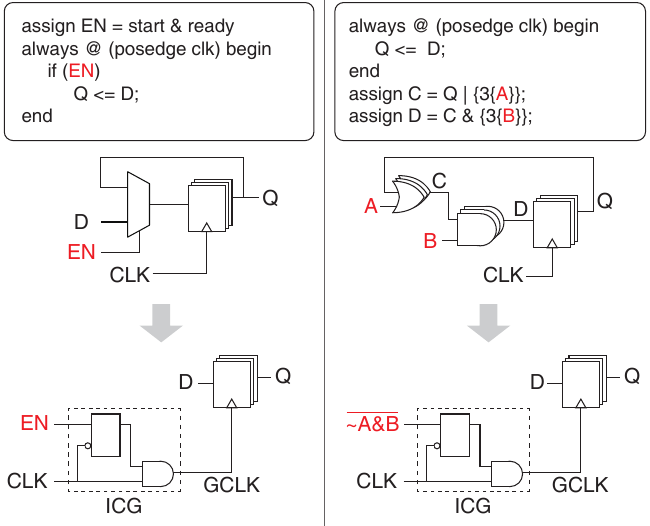}
    \caption{Clock-gating inference: (a) load-enable register and (b) combinational loop.}

    % with clock-gating cell (ICG cell) automatically inserted by synthesis tool. 
    \label{fig1}
    \vspace{-3mm}
\end{figure}

RTL clock gating is a popular method to reduce clock power consumption through RTL design. 
% It often provides the most significant reduction in power from a number of low-power techniques.
A gating condition is written in RTL code, so that synthesis tools can infer integrated clock-gating (ICG) cells from it.
\Cref{fig1}(a) illustrates a load-enable register, a frequent target of clock gating; it can easily be identified from the Verilog code and is implemented through a feedback multiplexer or by using an ICG cell if clock gating is inferred.
Another example of RTL clock gating is a combinational loop shown in \Cref{fig1}(b), where the gating condition is extracted whereas the condition is explicit in the example of \Cref{fig1}(a).

After logic synthesis with clock-gating inference, the efficiency of clock gating can be assessed through a number of measures \cite{ps}. Clock-gating ratio (CGR) is popular; it represents the percentage of flops connected to ICG cells or, equivalently, the percentage of flops whose clock can be gated. Flop-gating efficiency (FGE) denotes the percentage of cycles when a particular flop is clock-gated under a given switching activity, represented by FSDB or VCD.

%┌─────────────────────────────────────────────────┐

%┌─────────────────────────────────────────────────┐
%┌─────────────────────────────────────────────────┐
\begin{figure*}[t]
    \centering
    \includegraphics[scale=0.8]{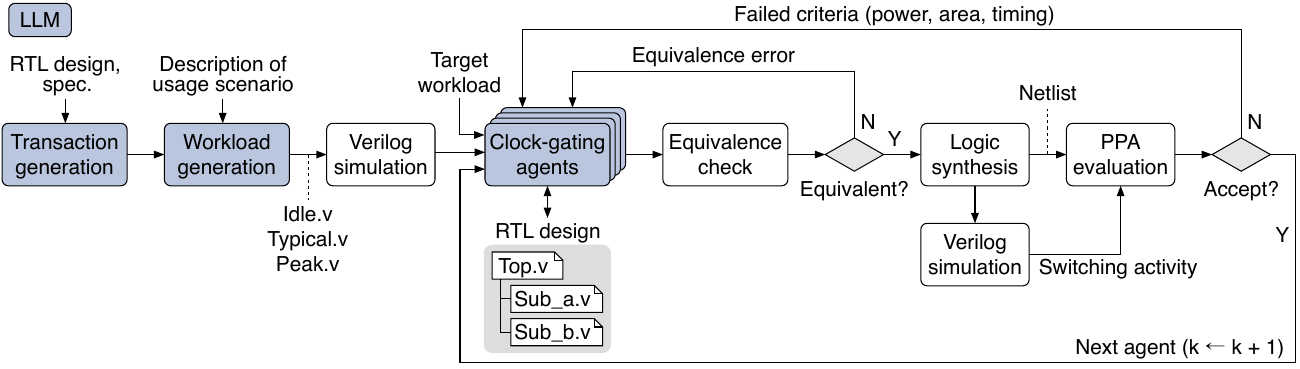}
    \caption{{Proposed flow for RTL clock gating.}}
    \label{fig2}
\end{figure*}
%┌─────────────────────────────────────────────────┐

Clock-gating efficiency (CGE), either CGR or FGE or both, is often not high enough \cite{low-cge2}, in particular at the early stage of RTL design. 
To improve CGE at the RTL level, one of the popular options that commercial tools take \cite{pa, joules, powerpro} is observability don't care (ODC) \cite{odc}. 
ODC identifies when a register's output is not utilized by downstream logic, which becomes a clock-gating condition. 
However, many other gating opportunities are missed, especially when a register's output is observable but kept unchanged.
Commercial tools can identify such registers, but they do not support automatic RTL correction.

\subsection{Motivations}

The actual RTL modification to improve CGE still relies on the designer \cite{thesis}. Developing effective gating functions requires an iterative process of code update and CGE analysis, which is time-consuming and demands a substantial understanding of both circuit functionality and switching activity. 

Large language models (LLMs) offer a promising alternative, as they can comprehend the functionality of Verilog code and modify it accordingly. Applying LLMs to RTL clock gating, however, poses two challenges. First, appropriate switching activity for FGE and power analysis is often unavailable, leaving LLMs unable to identify clock-gating opportunities. Although functional verification testbenches provide a convenient source of switching activity, they aim to cover as much design functionality as possible rather than reflect how the design is actually used. Second, LLMs lack expertise in RTL clock gating; as demonstrated in \Cref{sec3.B,sec3.C}, they fail to apply clock-gating techniques effectively without explicit guidance or when all techniques are provided at once.

This paper proposes an LLM-assisted refinement of Verilog code to improve CGE. LLMs first generate workloads representing specific usage scenarios \cite{usage}, i.e., Idle, Typical, and Peak, so that the refinement is guided by switching activity under the intended scenario. The refinement is then performed by four LLM clock-gating agents, each specialized in a single clock-gating technique.

\subsection{Proposed Method}

As illustrated in \Cref{fig2}, given the RTL design and its specification, an LLM generates transactions, each representing one type of operation of the design (e.g., encrypting a data block in AES), and combines them into three workloads representing usage scenarios (Idle, Typical, and Peak). Each workload is implemented as a Verilog testbench (or a C program for processors), whose simulation yields switching activity for each scenario. The switching activity of a user-selected target workload is used to evaluate FGE and power during RTL refinement.

The RTL design is then refined by four LLM-based clock-gating agents, each responsible for a different clock-gating technique. The agents operate sequentially, and each agent starts from the most recently accepted RTL. To help identify clock-gating opportunities, each agent is given a guideline for its technique and the 20 register groups with the highest waste scores under the target workload. Since practical designs are large and hierarchical, each agent iteratively invokes tools to read and modify Verilog code and to query the switching activity and FGE of selected registers. Once the agent has applied its technique to all applicable parts of the design, it submits the modified RTL for verification.

The submitted RTL is checked for functional equivalence against the original RTL. Once equivalence is verified, the RTL is synthesized, and the netlist is simulated under the target workload to evaluate power and FGE. The refined design is accepted only if it reduces power and satisfies the area and timing constraints. Any equivalence or PPA check failure triggers a feedback loop, providing the agent with error messages or evaluation reports to guide correction. If the RTL is not accepted within ten submissions, the agent's changes are discarded. This process is repeated for all agents.

%┌─────────────────────────────────────────────────┐
\begin{figure}[t]
    \centering
    \includegraphics[scale=0.8]{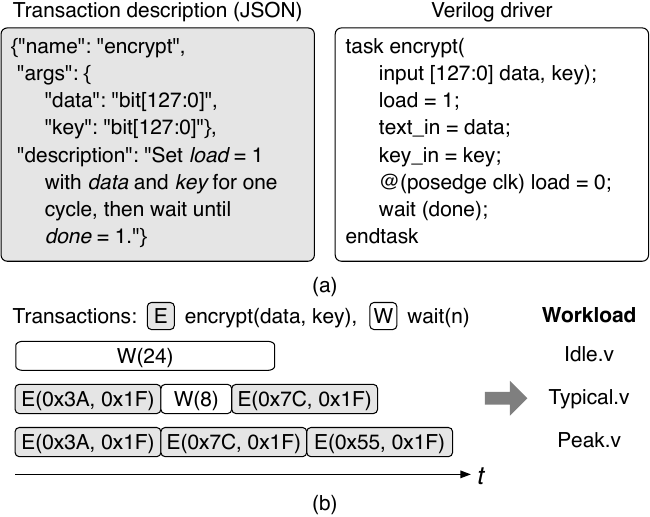}
    \caption{Workload generation for the AES encoder: (a) a transaction described in JSON and its Verilog driver, and (b) workloads composed of transactions.}
    \label{fig3}
    \vspace{-3mm}
\end{figure}
%┌─────────────────────────────────────────────────┐

\section{LLM-Assisted RTL Clock Gating}\label{sec2}

\subsection{Switching Activity Generation}\label{sec2.A}
As illustrated in \Cref{fig3}(a), each transaction is described in JSON, specifying its name, arguments with their types and bit-widths, and a natural-language description of input port signals for each clock cycle. For example, the AES encoder has two transactions: \texttt{encrypt(data, key)} for encrypting a data block and \texttt{wait(n)} for keeping the inputs inactive for a specified number of cycles. For each transaction, the LLM then generates a Verilog driver that converts its arguments into the corresponding input port values for each clock cycle. For processors, it instead generates a C function executed by the core.

The generated transactions are then used to construct workloads for three usage scenarios (Idle, Typical, and Peak), defined by how the \texttt{wait} transaction, common to all designs, is used. Idle consists only of \texttt{wait}, Typical interleaves \texttt{wait} with other transactions to represent typical usage, and Peak executes transactions back-to-back without \texttt{wait} at the fastest rate permitted by the design. Given these descriptions and the available transactions, the LLM generates a workload for each scenario by determining the order, arguments, and number of repetitions of transactions. \Cref{fig3}(b) illustrates the three workloads generated for the AES encoder. Each workload is then written as a Verilog file that executes its transaction sequence using the corresponding drivers, or as a C program for processors. Simulating the RTL design with the three workloads yields switching activity under each usage scenario.

%┌───────────────────────────────────────────────
\begin{figure}[t]
    \centering
    \includegraphics[scale=0.8]{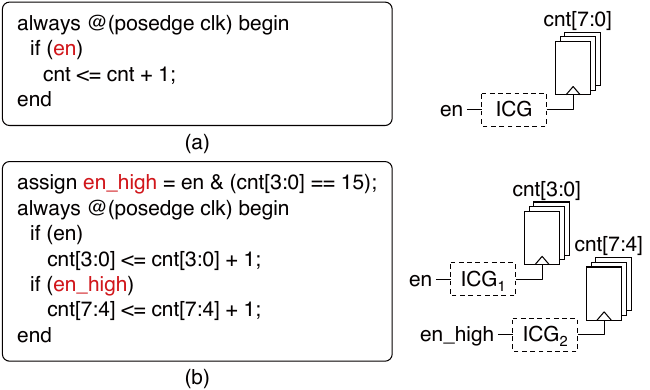}
    \vspace{-5mm}
    \caption{Clock gating of binary counter: (a) the whole register is gated and (b) the register is divided into two smaller ones, each gated separately.}
    \label{fig4}
    \vspace{-3mm}
\end{figure}
%┌───────────────────────────────────────────────

%┌───────────────────────────────────────────────

\subsection{LLM-Agents for Clock Gating}\label{sec2.B}

Once the user selects the target workload used to measure CGE and power, the Verilog code is refined. Since clock gating involves various implementation techniques, multiple agents are employed, each focusing on one technique. Each agent examines whether its technique is applicable and refines the code accordingly.

\subsubsection{Agent Role Design}\label{sec2.B.1}

Agent-1 aims to transform the non-load-enable registers, which are not clock-gated, into load-enable ones. Assume a register $data\_tmp$ with input signal $data\_in$; assume also that $data\_in$ is conditionally valid. The goal of this agent is to extract such condition, and convert $data\_tmp$ into a load-enable register while the condition is used for the gating function.

Agent-2 tries to refine the enable condition of load-enable register. Note from \Cref{fig1}(a) that the clock can be gated (EN = 0) more often if EN becomes more restrictive and becomes 1 less frequently. This is possible if EN is ANDed with additional conditions under which the register retains its current value. 

Agent-3 targets registers whose bits exhibit non-uniform switching activities and divides them into smaller registers, each containing bits with similar switching activities and gated separately. An example is illustrated in \Cref{fig4}(a) for an 8-bit binary counter ($cnt[7:0]$). Using a single gating function for $cnt$ causes unnecessary clock activity in the upper bits, because the lower bits toggle frequently while the upper bits toggle much less. Agent-3 may propose separate gating functions for $cnt[7:4]$ and $cnt[3:0]$, so that $cnt[7:4]$ is loaded only when $cnt[3:0]$ contains 4'b1111, as shown in \Cref{fig4}(b).

\begin{figure}[t]
    \centering
    \includegraphics[scale=0.8]{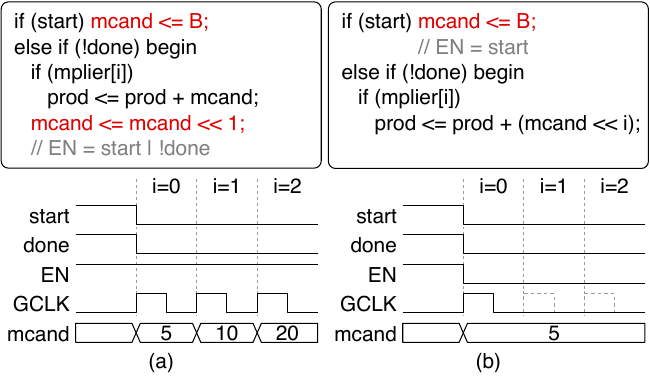}
    \vspace{-5mm}
\caption{Change in the gating condition for \texttt{mcand} through datapath restructuring in shift-and-add multiplication: (a) \texttt{start | !done} and (b) \texttt{start}.}
    \label{fig5}
\vspace{-3mm}
\end{figure}

Agent-4 modifies datapath computations to eliminate unnecessary register updates and improve clock gating. \Cref{fig5}(a) shows an example of shift-and-add multiplication, where $mplier$ is a multiplier and $mcand$ is a multiplicand. Register $mcand$ is loaded in two branches, so its enable condition is the OR of the two corresponding branch conditions. In \Cref{fig5}(b), $mcand$ is not loaded in the second branch; instead, it is shifted by $i$ positions when added to the partial product. Its enable condition then follows only the first branch and becomes more restrictive, allowing its clock to be gated more often.

%┌───────────────────────────────────────────────
%┌───────────────────────────────────────────────

\subsubsection{Agent Prompt and Tool Design}\label{sec2.B.2}

Each agent prompt consists of a system prompt and a user prompt. The system prompt includes (1) a clock-gating guideline describing the agent's technique with Verilog examples, (2) rules such as preserving functional correctness and satisfying area and timing constraints, and (3) descriptions of the available tools. The user prompt consists of (1) an instruction to reduce power consumption through RTL clock gating and (2) a register activity table under the target workload.

The activity table lists the register groups most likely to benefit from clock gating, ranked by a waste score. The score favors wide registers clocked far more often than their values change, and is defined as $W = B\max(0,(1-\mathrm{FGE})-U)$, where $B$ is the total number of flip-flop bits in a register group and $U$ is the fraction of cycles in which the register value changes under the target workload; thus, $(1-\mathrm{FGE})-U$ approximates the fraction of cycles in which the register is clocked without changing its value. The 20 register groups with the highest waste scores are included in the table, along with their module paths, bit widths, switching activity, and FGE.

Each agent iteratively refines the RTL design by invoking tools to inspect and edit Verilog files and to query detailed switching activity and FGE of individual registers, including those omitted from the activity table. Once the agent completes its refinement, it submits the modified RTL for verification and evaluation, as described in \Cref{sec2.C}.

\begin{comment}
    
Each agent prompt consists of a system prompt and user prompt. System prompt consists of (1) the technique description and (2) its guidelines.
System prompt에서 technique을 설명할 때에는 간단한 베릴로그 코드 예시를 함께 제공하였고, 새로운 gating function을 제안할 때 ~~~~ 를 ~~하도록 해서 새로운 gating function이 functionally correct하고 area overhead를 최소화할 수 있도록 유도하였다.

A user prompt consists of (1) the instruction to apply the technique, (2) the list of target registers, (3) the corresponding switching context, (4) the Verilog code, and (5) the fixed action format.
As shown in \Cref{fig3}, agent별로 서로 다른 register 목록과 switching context가 제공해 해당 기법을 ~~ 더 잘 탐색할 수 있도록 하였다.
Agent-1에게는 non-load-enable-register의 toggle rate을 제공하고, Agent-2의 경우 load-enable-register의 toggle rate을 제공하며, 추가로 enable signal의 duty cycle까지 함께 제공해 현재 gating function이 얼마나 자주 clock을 gating하고 있는지 쉽게 판단할 수 있게 한다.
Agent-3의 경우 register 하나에서 bit 별 switching activity를 포착할 수 있어야 하기 때문에 bit-level toggle rate을 제공한다.
마지막으로 Agent-4는 sub-module의 gating 가능 여부를 판단할 수 있어야 하기 때문에 sub-module state transition 값을 제공해준다.

The fixed output format enforces LLMs to generate intermediate reasoning steps, allowing them to derive more accurate and efficient gating functions based on the chain-of-thought (CoT) principle. The action includes steps for (1) current limitation in clock-gating efficiency in natural language,(analysis of target register or sub-module), (2) applicability of its technique for each register, and (3) a new gating function in Verilog code.
\end{comment}

\subsection{RTL Verification and Evaluation}\label{sec2.C}

Sequential equivalence checking via \cite{slec} verifies the submitted RTL against the original design. If the check fails, the error messages are fed back to the agent, along with the previous prompt-response interaction, for revision.

The verified RTL is then synthesized and simulated with the testbench for the target workload (\Cref{sec2.A}) to evaluate power and FGE. It is accepted only if (1) its power is lower than that of the previously accepted design, (2) its area overhead relative to the original design does not exceed 10\%, and (3) it meets timing at the target clock period after synthesis (WNS $\geq 0$). Otherwise, the design is restored to the previously accepted one, and the agent receives the rejection reason and evaluation reports. If no submission is accepted within ten attempts, the agent's changes are discarded.

Once an agent finishes, the next agent starts from the most recently accepted RTL. The process continues until all four agents have completed their refinement, and the last accepted RTL becomes the final design.

%______________________
\begin{table}[t]
\centering
\caption{Benchmark designs.}
\label{tab1}
\footnotesize
\setlength{\tabcolsep}{2.5pt}
\setlength{\aboverulesep}{0pt}
\setlength{\belowrulesep}{0pt}
\renewcommand{\arraystretch}{1.1}
\setlength{\extrarowheight}{1pt}
\begin{tabular}{@{}l|ccccc@{}}
\toprule
 & \#Code & & & & CGR \\[-1.5pt]
Function & lines & \#Modules & \#FFs & \#Cells & (\%) \\
\midrule
AES-128 encryption \cite{realbench} & 471 & 4 & 530 & 9,055 & 24.9 \\
AES-128 decryption \cite{realbench} & 793 & 5 & 2,072 & 12,807 & 80.7 \\
UART \cite{ours}        & 1,648 & 10 & 604 & 2,004 & 80.1 \\
SD card controller \cite{realbench} & 2,751 & 14 & 2,643 & 6,182 & 94.5 \\
E203 RISC-V core \cite{realbench}   & 13,668 & 66 & 2,319 & 14,977 & 96.8 \\
\bottomrule
\end{tabular}
\end{table}
%______________________

%------------------------------------------------------
% Section 4. Experimental Results
%------------------------------------------------------
\section{Experimental Results}\label{sec4}

We evaluate five RTL designs listed in \Cref{tab1}: AES encoder, AES decoder, UART, SD card controller, and RISC-V core. Four designs are taken from \cite{realbench}, and UART from \cite{ours}. The designs are not modified to expose clock-gating opportunities, except for minor fixes for tool compatibility.

The entire flow in \Cref{fig2} is automated using Python scripts, with Claude Sonnet 5.5 \cite{claude5.5} (medium reasoning effort) as the LLM. Verilog simulation \cite{vcs} is performed with each LLM-generated workload to generate an FSDB, from which the FGE of the RTL design is estimated using \cite{ps}. Each RTL design is then synthesized using \cite{dc} with a 14-nm standard-cell library \cite{saed14}, applying \texttt{compile\_ultra -gate\_clock} and \texttt{set\_clock\_gating\_style -minimum\_bitwidth 3}, with the target clock period set to 1.2 times the critical path delay of the original design. The netlist is simulated again with the same workload, and the resulting FSDB is used to measure FGE using \cite{pt} and, after conversion to SAIF, to estimate dynamic power using \cite{pp}. For the RISC-V core, the power of the SRAM, modeled using \cite{fakeram}, is excluded to focus on logic power.

%__________________________________
\begin{table}[t]
\centering
\caption{Workloads generated by the LLM.}
\label{tab2}
\footnotesize
\setlength{\tabcolsep}{1.8pt}
\setlength{\aboverulesep}{0pt}
\setlength{\belowrulesep}{0pt}
\renewcommand{\arraystretch}{1.1}
\setlength{\extrarowheight}{1pt}
\ifdefined\covw\else\newlength{\covw}\fi
\settowidth{\covw}{Code coverage}
\setlength{\covw}{\dimexpr(\covw-2\tabcolsep)/2\relax}% Line and Toggle share the header width evenly
\begin{tabular}{@{}c|l|cc|cc|w{c}{\covw}w{c}{\covw}@{}}
\toprule
 & & \multicolumn{2}{c|}{Transactions} & \multicolumn{2}{c|}{Simulated cycles} & \multicolumn{2}{c}{Code coverage} \\[-1.5pt]
\cmidrule(lr){3-4} \cmidrule(lr){5-6} \cmidrule(lr){7-8}
Design & Workload & {\renewcommand{\arraystretch}{0.9}\setlength{\extrarowheight}{0pt}\begin{tabular}[b]{@{}c@{}}\rule{0pt}{9pt}Types\\(\#)\end{tabular}} & {\renewcommand{\arraystretch}{0.9}\setlength{\extrarowheight}{0pt}\begin{tabular}[b]{@{}c@{}}\rule{0pt}{9pt}Total\\(\#)\end{tabular}} & {\renewcommand{\arraystretch}{0.9}\setlength{\extrarowheight}{0pt}\begin{tabular}[b]{@{}c@{}}\rule{0pt}{9pt}Total\\(K)\end{tabular}} & {\renewcommand{\arraystretch}{0.9}\setlength{\extrarowheight}{0pt}\begin{tabular}[b]{@{}c@{}}\rule{0pt}{9pt}Active\\(\%)\end{tabular}} & {\renewcommand{\arraystretch}{0.9}\setlength{\extrarowheight}{0pt}\begin{tabular}[b]{@{}c@{}}\rule{0pt}{9pt}Line\\(\%)\end{tabular}} & {\renewcommand{\arraystretch}{0.9}\setlength{\extrarowheight}{0pt}\begin{tabular}[b]{@{}c@{}}\rule{0pt}{9pt}Toggle\\(\%)\end{tabular}} \\
\midrule
\multirow{4}{*}{\renewcommand{\arraystretch}{0.95}\setlength{\extrarowheight}{0pt}\begin{tabular}[c]{@{}c@{}}AES\\encoder\end{tabular}} & Idle & 1 & 1 & 0.2 & 0 & 55 & 73 \\
 & Typical & 2 & 22 & 0.4 & 38 & 71 & 89 \\
 & Peak & 1 & 16 & 0.2 & 100 & 45 & 86 \\
\cline{2-8}
 & All & 2 & 39 & 0.8 & 45 & 90 & 89 \\
\midrule
\multirow{4}{*}{\renewcommand{\arraystretch}{0.95}\setlength{\extrarowheight}{0pt}\begin{tabular}[c]{@{}c@{}}AES\\decoder\end{tabular}} & Idle & 1 & 1 & 0.1 & 0 & 45 & 69 \\
 & Typical & 3 & 18 & 0.3 & 46 & 74 & 89 \\
 & Peak & 1 & 16 & 0.2 & 100 & 56 & 86 \\
\cline{2-8}
 & All & 3 & 35 & 0.7 & 52 & 93 & 90 \\
\midrule
\multirow{4}{*}{\renewcommand{\arraystretch}{0.95}\setlength{\extrarowheight}{0pt}\begin{tabular}[c]{@{}c@{}}UART\end{tabular}} & Idle & 1 & 1 & 5.0 & 0 & 29 & 1 \\
 & Typical & 3 & 24 & 4.4 & 43 & 52 & 38 \\
 & Peak & 2 & 24 & 3.7 & 97 & 52 & 39 \\
\cline{2-8}
 & All & 3 & 49 & 13.1 & 42 & 52 & 41 \\
\midrule
\multirow{4}{*}{\renewcommand{\arraystretch}{0.95}\setlength{\extrarowheight}{0pt}\begin{tabular}[c]{@{}c@{}}SD card\\controller\end{tabular}} & Idle & 1 & 1 & 100.0 & 0 & 16 & 1 \\
 & Typical & 5 & 21 & 27.8 & 63 & 70 & 48 \\
 & Peak & 2 & 11 & 26.5 & 98 & 70 & 47 \\
\cline{2-8}
 & All & 5 & 33 & 154.4 & 28 & 70 & 49 \\
\midrule
\multirow{4}{*}{\renewcommand{\arraystretch}{0.95}\setlength{\extrarowheight}{0pt}\begin{tabular}[c]{@{}c@{}}RISC-V\\core\end{tabular}} & Idle & 1 & 1 & 20.0 & 0 & 25 & 0 \\
 & Typical & 4 & 23 & 89.0 & 51 & 74 & 55 \\
 & Peak & 1 & 1 & 5.6 & 100 & 62 & 38 \\
\cline{2-8}
 & All & 5 & 25 & 114.5 & 44 & 74 & 56 \\

\bottomrule
\end{tabular}
\end{table}
%__________________________________

% ---------------- Table 3: results under the target workload (two columns wide) ----------------
\begin{table*}[t]
\caption{Post-CTS CGR, FGE, power, area, and WNS under the target workload for three RTL refinement methods: commercial tool vs.\ LLM baseline vs.\ proposed method (averaged over three runs, except for the commercial tool, which is run once; WNS: worst of the three runs; Agent: clock-gating agents whose refinements are accepted)}
\label{tab3}
\centering\scriptsize
\setlength{\tabcolsep}{2pt}
\renewcommand{\arraystretch}{1.1}
\setlength{\extrarowheight}{1pt}
\setlength{\aboverulesep}{0pt}
\setlength{\belowrulesep}{0pt}
\begin{tabular}{@{}c|l|ccccc|rrrrcc|rrrrcc|rrrrc|cc@{}}
\toprule
\multirow{3}{*}{Design} & \multirow{3}{*}{\makecell[l]{Target\\workload}}
 & \multicolumn{5}{c|}{\multirow{2}{*}{Original}} & \multicolumn{19}{c}{After refinement} \\
\cmidrule(lr){8-26}
 & & \multicolumn{5}{c|}{} & \multicolumn{6}{c|}{Commercial tool~\cite{powerpro}} & \multicolumn{6}{c|}{LLM baseline}
 & \multicolumn{7}{c}{Multi-agent (Proposed method)} \\
\cmidrule(lr){3-7}\cmidrule(lr){8-13}\cmidrule(lr){14-19}\cmidrule(lr){20-26}
 & & \makecell{CGR\\(\%)} & \makecell{FGE\\(\%)} & \makecell{Pwr\\($\mu$W)} & \makecell{Area\\($\mu$m$^2$)} & \makecell{WNS\\(ps)}
 & \hv{\makecell{$\Delta$CGR\\(\%p)}} & \hv{\makecell{$\Delta$FGE\\(\%p)}} & \hv{\makecell{$\Delta$Pwr\\(\%)}} & \hv{\makecell{$\Delta$Area\\(\%)}}
 & \makecell{WNS\\(ps)} & \makecell{Time\\(min)}
 & \hv{\makecell{$\Delta$CGR\\(\%p)}} & \hv{\makecell{$\Delta$FGE\\(\%p)}} & \hv{\makecell{$\Delta$Pwr\\(\%)}} & \hv{\makecell{$\Delta$Area\\(\%)}}
 & \makecell{WNS\\(ps)} & \makecell{Time\\(min)}
 & \hv{\makecell{$\Delta$CGR\\(\%p)}} & \hv{\makecell{$\Delta$FGE\\(\%p)}} & \hv{\makecell{$\Delta$Pwr\\(\%)}} & \hv{\makecell{$\Delta$Area\\(\%)}}
 & \makecell{WNS\\(ps)} & Agent & \makecell{Time\\(min)} \\
\midrule
% Columns after Original: Commercial tool (dCGR dFGE dPwr dArea WNS Runtime) | LLM baseline (same) | Proposed (dCGR dFGE dPwr dArea WNS Agent Runtime)
\mr{\makecell{AES\\encoder}} & Idle    & \mr{25} & 25 & 3,722 & \mr{3,113} & \mr{57}
  & 0 & 0 & 0 & 0.0 & 57 & 1 & 0 & 0 & 0 & 0.0 & 57 & 0 & +48 & +48 & $-$59 & +0.5 & 47 & 1 & 84 \\
              & Typical &         & 24 & 3,761 &            &
  & 0 & 0 & 0 & 0.0 & 57 & 1 & 0 & 0 & 0 & 0.0 & 57 & 0 & +48 & +40 & $-$50 & $-$0.1 & 42 & 1 & 34 \\
              & Peak    &         & 22 & 3,811 &            &
  & 0 & 0 & 0 & 0.0 & 57 & 1 & 0 & 0 & 0 & 0.0 & 57 & 0 & +48 & +28 & $-$23 & +0.2 & 58 & 1, 4 & 28 \\
\midrule
\mr{\makecell{AES\\decoder}} & Idle    & \mr{81} & 75 & 3,702 & \mr{5,270} & \mr{73}
  & 0 & 0 & 0 & 0.0 & 73 & 5 & +7 & +7 & $-$35 & $-$0.3 & 23 & 6 & +7 & +13 & $-$38 & $-$0.1 & 21 & 1, 2 & 14 \\
              & Typical &         & 75 & 3,840 &            &
  & 0 & 0 & 0 & 0.0 & 73 & 5 & +7 & +7 & $-$30 & $-$0.3 & 23 & 6 & +7 & +10 & $-$31 & +0.1 & 43 & 1, 2 & 12 \\
              & Peak    &         & 75 & 4,189 &            &
  & 0 & 0 & 0 & 0.0 & 73 & 1 & +7 & +7 & $-$30 & $-$0.3 & 23 & 6 & +7 & +7 & $-$28 & $-$0.2 & 18 & 1 & 7 \\
\midrule
\mr{UART}     & Idle    & \mr{80} & 73 & 600 & \mr{1,094} & \mr{142}
  & +7 & +14 & $-$27 & +6.8 & 156 & 4 & +5 & +5 & $-$20 & +0.6 & 138 & 9 & +7 & +7 & $-$26 & +0.1 & 154 & 1--3 & 30 \\
              & Typical &         & 73 & 614 &            &
  & +2 & +7 & $-$19 & +5.5 & 147 & 4 & 0 & 0 & 0 & 0.0 & 142 & 6 & +9 & +10 & $-$41 & +0.3 & 156 & 1--4 & 27 \\
              & Peak    &         & 72 & 636 &            &
  & +7 & +9 & $-$16 & +4.3 & 149 & 4 & +5 & +3 & $-$14 & +0.6 & 138 & 10 & +7 & +7 & $-$19 & +2.9 & 170 & 1--4 & 30 \\
\midrule
\mr{\makecell{SD card\\controller}} & Idle    & \mr{95} & 85 & 1,073 & \mr{3,907} & \mr{140}
  & 0 & 0 & 0 & 0.0 & 140 & 5 & 0 & 0 & 0 & 0.0 & 140 & 5 & 0 & +4 & $-$16 & +0.4 & $-$4 & 1, 2, 4 & 48 \\
              & Typical &         & 85 & 1,089 &            &
  & 0 & 0 & 0 & 0.0 & 140 & 6 & 0 & 0 & 0 & 0.0 & 140 & 6 & 0 & +2 & $-$3 & 0.0 & 45 & 3 & 31 \\
              & Peak    &         & 85 & 1,093 &            &
  & 0 & 0 & 0 & 0.0 & 140 & 6 & 0 & 0 & 0 & 0.0 & 140 & 6 & 0 & +2 & $-$6 & +0.5 & 91 & 2 & 24 \\
\midrule
\mr{\makecell{RISC-V\\core}} & Idle    & \mr{97} & 97 & 26 & \mr{50,048} & \mr{1,792}
  & 0 & 0 & 0 & 0.0 & 1,792 & 2 & 0 & 0 & 0 & 0.0 & 1,792 & 0 & 0 & +1 & $-$6 & 0.0 & 1,855 & 1 & 20 \\
              & Typical &         & 92 & 99 &             &
  & 0 & 0 & 0 & 0.0 & 1,792 & 2 & 0 & 0 & 0 & 0.0 & 1,792 & 1 & 0 & +2 & $-$3 & 0.0 & 1,653 & 1, 2, 4 & 54 \\
              & Peak    &         & 88 & 192 &            &
  & 0 & 0 & 0 & 0.0 & 1,792 & 4 & 0 & 0 & 0 & 0.0 & 1,792 & 0 & 0 & +2 & $-$4 & 0.0 & 1,834 & 1 & 29 \\
\bottomrule
\end{tabular}
\end{table*}

\subsection{LLM-Generated Workloads}\label{sec3.A}

For each design, the LLM generates Idle, Typical, and Peak workloads as described in \Cref{sec2.A}. To verify that each workload represents its intended usage scenario (Idle issues no requests, Typical interleaves requests with idle intervals, and Peak issues requests back-to-back), \Cref{tab2} reports the number of transaction types, the total number of transactions, the number of simulated cycles, the fraction of active cycles, and code coverage of each workload. An active cycle is a cycle in which the design is processing a request, from its acceptance to its completion. For the RISC-V core, active cycles are measured as cycles outside WFI-based sleep periods.

For the non-processor designs, the LLM composes the Idle workload using only the \texttt{wait} transaction, whereas it uses the largest number of transaction types (2--5) for the Typical workload. For the Peak workload, it uses fewer types (1--2), repeating only the main operations. For the RISC-V core, the LLM generates a C program whose transactions are C functions, including control-oriented routines (e.g., sorting) and an ALU-intensive kernel (e.g., matrix multiplication), while \texttt{wait} is implemented as WFI-based sleep.

Active cycles account for 0\%, 38--63\%, and 97--100\% of the Idle, Typical, and Peak workloads across the five designs, showing that the generated workloads reflect the distinct activity levels of the three scenarios. For Peak, the LLM also selects argument values that maximize the operating rate (e.g., a baud divisor of 1 for UART). Combining the workloads of all scenarios, the five designs achieve 52--93\% line coverage, excluding the reset and configuration phases. The lowest value is for UART, whose reset and configuration code is executed only outside the measured workloads; including the entire simulation raises its coverage to 82\%.

\subsection{FGE Improvement and Power Reduction}\label{sec3.B}

The proposed method (\Cref{sec2.B}) refines the RTL with four agents, each specialized in one clock-gating technique, using guidelines and the 20 most promising candidates ranked by waste score (\Cref{sec2.B.2}). We compare it with two baselines: a commercial tool \cite{powerpro} that automatically applies stability-based clock gating (gating a register when its input remains unchanged) and ODC-based clock gating (when its output is not observed), and an LLM baseline in which a single agent receives the same candidates but no technique-specific guidelines, with all other settings unchanged. The two LLM-based methods are run three times per target workload and averaged, whereas the commercial tool is run once. Each refined design is implemented through CTS \cite{icc2} and evaluated under its target workload.

As shown in \Cref{tab3}, the proposed method improves FGE and reduces power for all designs and target workloads, by 23.6\% on average and up to 59\%, with an area overhead of at most 2.9\%; WNS after CTS is negative only for the SD card controller refined for Idle ($-$4~ps). In contrast, the commercial tool succeeds only for UART, and the LLM baseline only for the AES decoder and UART, reducing power by 4.1\% and 8.6\% on average, respectively.

Even for UART, the commercial tool generates gating conditions by adding comparators and flip-flops that detect whether a register input remains constant; the 32-bit register that latches bus write data is gated only when its input remains zero for two consecutive cycles. For the same register, both LLM-based methods derive the gating condition from the existing bus control signals \texttt{wb\_cyc\_i} and \texttt{wb\_stb\_i}, which indicate when the input is valid, resulting in smaller area overheads (at most 0.6\% and 2.9\% vs.\ 4.3--6.8\%). The proposed method also achieves comparable or larger power reductions (26\%, 41\%, and 19\% vs.\ 27\%, 19\%, and 16\% under Idle, Typical, and Peak).

Although the LLM baseline receives the same candidates, all of its refinements are limited to converting registers into load-enable registers, the most common technique (Agent-1). For the RISC-V core, for example, it concludes that the cycle counter cannot be gated because its value changes every cycle, whereas the proposed method splits the counter and gates its upper bits, which toggle much less often than the lower bits. As shown in \Cref{tab3}, the proposed method applies various techniques across the designs, suggesting that, without explicit guidelines, an LLM rarely explores techniques beyond adding an enable.

All designs refined by the proposed method are functionally equivalent to the originals, whereas some refinements by the commercial tool, e.g., for the AES decoder, are rejected because their equivalence cannot be verified. However, equivalence checking and synthesis dominate the runtime, which reaches up to 84 minutes.

% ---------------- Table 5: register-level FGE improvement per target, one column ----------------
\begin{table}[t]
\caption{Register-level FGE improvement under different refinement targets: original FGE (\%) under Idle/Typical/Peak (I/T/P) and whether FGE improved for each target workload (\checkmark: improvement in the majority of runs)}
\label{tab4}
\centering\footnotesize
\setlength{\tabcolsep}{3pt}
\renewcommand{\arraystretch}{1.1}
\setlength{\extrarowheight}{1pt}
\setlength{\aboverulesep}{0pt}
\setlength{\belowrulesep}{0pt}
\ifdefined\inpw\else\newlength{\inpw}\fi
\settowidth{\inpw}{100}% all six I/N/P columns share the width of "100"
\begin{tabular}{c|l|w{c}{\inpw}w{c}{\inpw}w{c}{\inpw}|w{c}{\inpw}w{c}{\inpw}w{c}{\inpw}}
\toprule
\multirow{2}{*}{Design} & \multirow{2}{*}{Register (\#FFs)}
 & \multicolumn{3}{c|}{\makecell{Original\\FGE (\%)}}
 & \multicolumn{3}{c}{\makecell{FGE improved\\under target}} \\
\cmidrule(lr){3-5}\cmidrule(lr){6-8}
 & & \multicolumn{1}{c}{I} & \multicolumn{1}{c}{T} & \multicolumn{1}{c|}{P} & I & T & P \\
\midrule
\multirow{2}{*}{AES enc.} & State (16$\times$8)          & 0 & 0 & 0 & \y & \y & \y \\
                          & Output text (128)           & 0 & 0 & 0 & \y & \y & \y \\
\midrule
\multirow{2}{*}{AES dec.} & Key expansion (140)         & 0 & 0 & 0 & \y & \y & \y \\
                          & Round key (4$\times$32)     & 0 & 7 & 15 & \y & \y & \n \\
\midrule
\multirow{2}{*}{UART}     & Bus write data (32)         & 0 & 0 & 0 & \y & \y & \y \\
                          & Divisor counter (16)        & 0 & 0 & 0 & \n & \y & \n \\
\midrule
\multirow{4}{*}{SD ctrl.} & Cmd.\ output buffer (40)    & 0 & 0 & 0 & \y & \y & \y \\
                          & Cmd.\ CRC (8)               & 0 & 0 & 0 & \n & \y & \y \\
                          & Cmd.\ argument (32)         & 0 & 60 & 98 & \y & \n & \n \\
                          & DMA address (32)            & 0 & 60 & 98 & \y & \n & \n \\
\midrule
\multirow{2}{*}{RISC-V}   & Cycle counter (32)          & 0 & 0 & 0 & \y & \y & \y \\
                          & Retired-instr.\ counter (32) & 100 & 57 & 3 & \n & \y & \y \\
\bottomrule
\end{tabular}
\end{table}
% ________________________________________

% __________________ Table 4: cross evaluation (power), one column __________________
\begin{table}[t]
\caption{Post-CTS power changes (\%) of designs refined for different target workloads, evaluated under
all three workloads (gray: target = evaluation workload; bold: largest power reduction
for each evaluation workload)}
\label{tab5}
\centering\footnotesize
\setlength{\tabcolsep}{4pt}
\renewcommand{\arraystretch}{1.1}
\setlength{\extrarowheight}{1pt}
\setlength{\aboverulesep}{0pt}
\setlength{\belowrulesep}{0pt}
\ifdefined\evw\else\newlength{\evw}\fi
\settowidth{\evw}{Typical}% Idle, Typical, and Peak columns share the width of "Typical"
\begin{tabular}{c|l|w{r}{\evw}|w{r}{\evw}|w{r}{\evw}}
\toprule
\multirow{2}{*}{Design} & \multirow{2}{*}{\makecell[l]{Refinement\\target workload}} & \multicolumn{3}{c}{Evaluation workload} \\
\cmidrule(lr){3-5}
 & & \hv{Idle} & \hv{Typical} & \multicolumn{1}{c}{Peak} \\
\midrule
\mr{AES enc.} & Idle    & \tg$-$59.3 & $-$43.3 & $-$17.6 \\
              & Typical & \textbf{$-$68.4} & \tg$-$50.4 & $-$20.7 \\
              & Peak    & $-$67.7 & \textbf{$-$50.7} & \tg\textbf{$-$22.6} \\
\midrule
\mr{AES dec.} & Idle    & \tg$-$38.4 & $-$30.6 & \textbf{$-$29.1} \\
              & Typical & \textbf{$-$38.6} & \tg\textbf{$-$30.7} & \textbf{$-$29.1} \\
              & Peak    & $-$32.5 & $-$27.9 & \tg$-$28.1 \\
\midrule
\mr{UART}     & Idle    & \tg$-$25.9 & $-$25.2 & $-$20.1 \\
              & Typical & \textbf{$-$42.1} & \tg\textbf{$-$40.8} & \textbf{$-$36.1} \\
              & Peak    & $-$20.6 & $-$20.1 & \tg$-$19.3 \\
\midrule
\mr{SD ctrl.} & Idle    & \tg\textbf{$-$15.7} & \textbf{$-$7.7} & $-$2.8 \\
              & Typical & $-$3.8 & \tg$-$3.4 & $-$3.3 \\
              & Peak    & $-$6.5 & $-$6.2 & \tg\textbf{$-$6.0} \\
\midrule
\mr{RISC-V}   & Idle    & \tg$-$6.1 & $-$0.9 & 0.0 \\
              & Typical & $-$4.9 & \tg$-$3.1 & $-$2.6 \\
              & Peak    & \textbf{$-$12.5} & \textbf{$-$5.1} & \tg\textbf{$-$3.6} \\
\bottomrule
\end{tabular}
\end{table}

\subsection{Impact of Target Workload}
The proposed refinement is tailored to a target workload in two ways: the agents receive the 20 most promising candidates ranked by waste score under that workload (\Cref{sec2.B.2}), and a refined design is accepted only if it reduces power under the same workload (\Cref{sec2.C}). To examine the impact of this targeting, we also evaluate each design refined in \Cref{tab3} under the two non-target workloads. \Cref{tab4} lists all registers whose FGE improves under at least one target workload, along with their original FGE, and \Cref{tab5} reports the resulting power changes.

As shown in \Cref{tab4}, registers with zero original FGE under all workloads, such as the AES state and the RISC-V cycle counter, are gated regardless of the target. In contrast, registers whose FGE varies across workloads are gated only when the target workload is one under which their FGE is low. For example, the command argument and DMA address registers of the SD card controller have FGE of 0\%, 60\%, and 98\% under Idle, Typical, and Peak, and thus are gated only when targeting Idle. Conversely, the RISC-V retired-instruction counter is already fully gated under Idle (100\%) and is therefore gated only when targeting Typical or Peak.

As shown in \Cref{tab5}, every refined design reduces or maintains power under all workloads, since most gated registers have low FGE under all workloads. Thus, a design that gates registers missed by other targets can perform best even under non-target workloads: refining for the evaluation workload itself yields the largest reduction in only 6 of the 15 cases. For example, the SD card controller refined for Idle, the only one that gates the command argument and DMA address registers, performs best under Idle and Typical, and the UART refined for Typical does so under all three workloads. This suggests that, since each target exposes different opportunities (e.g., different counters in UART), combining candidates from multiple workloads may further improve refinement.

%The target workload affects which clock-gating opportunities the agent discovers. As shown in \Cref{tab3}, refining the same design for different target workloads yields different FGE improvements and power reductions. For example, the 128-bit key registers of the AES decoder undergo redundant updates during Idle periods, whereas their updates are primarily associated with active operations under Peak. The agent targeting Peak does not gate these registers, resulting in a 6.2 percentage point smaller FGE increase under Idle than the agent targeting Idle.

%The refined designs also reduce power under workloads other than their targets: all refined designs reduce or maintain power across all three workloads, as shown by each row of \Cref{tab3}. For example, the AES encoder refined for Idle reduces power by 59.3\%, 43.3\%, and 17.6\% under Idle, Typical, and Peak, respectively. However, targeting a particular workload does not always yield the largest power reduction under that workload, as the gray cells achieve the largest reduction in only 6 of the 15 evaluation cases.

\subsection{Ablation Study: Multi-Agent Refinement}\label{sec3.C}

To evaluate the effectiveness of multiple agents, we compare the proposed method with two single-agent variants, with all other settings unchanged: one can read each technique-specific guideline from a separate skill document on demand, and the other receives all four guidelines in its prompt. Each variant is run once for each of the 15 cases, i.e., the five designs refined for each of the three target workloads, and the results of the proposed method are taken from \Cref{tab3}.

As shown in \Cref{tab6}, the single agent with skill documents refines only 7 of the 15 cases, improving FGE by 2.1\%p and reducing power by 10.1\% on average. Although the agent is explicitly instructed on the availability of the skill documents and how to read them, it never opens them in any of the 15 runs. The single agent with all guidelines in its prompt refines 9 cases, improving FGE by 2.3\%p and reducing power by 8.5\%. It applies at most one technique per run, mostly load-enable conversion (Agent-1; 8 of the 15 runs) and rarely the others (Agent-2, -3, and -4 in 1, 3, and 0 runs, respectively). Both ignoring the skill documents and applying only one technique stem from LLM laziness \cite{lazy,lazy2}, where the model neglects some instructions. In contrast, the proposed method assigns a dedicated agent to each technique, so that all four techniques are explored and contribute to the improvement (\Cref{tab3}).

% ---------------- Ablation table: single agent vs. multi-agent (one column) ----------------
% Averages over the 15 design-workload cases. Single-agent values: one run each.
% Proposed values: from Table 3 (average of three runs).
\begin{table}[t]
\caption{Comparison of three agent configurations: single agent with skill documents vs.\ single
agent with guidelines in the prompt vs.\ multiple agents (proposed method)}
\label{tab6}
\centering\footnotesize
\setlength{\tabcolsep}{4pt}
\renewcommand{\arraystretch}{1.1}
\setlength{\extrarowheight}{1pt}
\setlength{\aboverulesep}{0pt}
\setlength{\belowrulesep}{0pt}
\begin{tabular}{@{}l|c|c|c@{}}
\toprule
 & \makecell{Single agent\\+ Skill} & \makecell{Single agent\\+ Prompt} & \makecell{Multi-agent\\+ Prompt (Proposed)} \\
\midrule
Average $\Delta$FGE (\%p)  & +2.1    & +2.3    & +12.1   \\
Average $\Delta$Power (\%) & $-$10.1 & $-$8.5  & $-$23.6 \\
Refined cases              & 7/15    & 9/15    & 15/15   \\
\bottomrule
\end{tabular}
\vspace{-3mm}

\end{table}

%\subsubsection{Effectiveness of Multiple Clock-Gating Agents}\label{sec3.C.2}
%To evaluate the effectiveness of multiple clock-gating agents, we compare the proposed method against two variants: (1) a no-guidance approach, where the LLM receives no clock-gating technique descriptions, and (2) a single-agent approach, where one LLM agent receives all four technique descriptions simultaneously. Other steps are kept the same. \Cref{tab5} compares the power reduction and runtime of the three approaches under the Idle workload after synthesis.

%As shown in \Cref{tab5}, the no-guidance approach reduces power in only two of the five designs, suggesting that LLMs struggle to identify effective clock-gating opportunities without explicit guidance. Providing all four techniques to a single agent improves some results but does not consistently outperform no guidance, as shown by AES decoder (0.0\% vs. 38.5\%). In contrast, our multi-agent approach achieves the largest power reduction across all five designs (15.3--55.7\%), indicating that guiding the LLM to apply one technique at a time is more effective than providing all techniques at once, although it increases runtime to 14--84 minutes.

\section{Related Works}

Recent studies have explored LLM-assisted RTL refinement by improving LLM capabilities or guiding their RTL modifications. One study fine-tunes an LLM through direct preference optimization using functionally equivalent RTL designs ranked by user-weighted PPA costs \cite{ppartl}, while another uses reinforcement learning with advantages computed from correctness checks and PPA ratios relative to a reference design \cite{chipseek}.

Other studies guide existing LLMs in RTL optimization. An LLM is given post-synthesis reports, PPA objectives, and optimization techniques (e.g., clock gating), from which it selects techniques \cite{veriopt}. Monte Carlo tree search is used to select retrieved optimization instructions, algorithms, and code examples for RTL rewriting \cite{rtlrewriter}, while retrieval-augmented generation is used to select optimization rules and generate AST-based transformations \cite{symrtlo}. Multiple agents also collaborate on planning, RTL rewriting, and evaluation \cite{codmas}.

For low-power RTL design, however, two aspects remain underexplored. First, prior work has not investigated how to generate switching activity reflecting actual usage scenarios for RTL refinement. Second, while existing approaches provide multiple optimization techniques together, how to effectively guide LLMs in applying individual clock-gating techniques remains unclear.

\section{Conclusion} \label{sec6}

We have presented LLM-assisted refinement of Verilog code to improve CGE. LLMs generate Idle, Typical, and Peak workloads to obtain switching activity under each usage scenario, and the registers with low FGE under the selected workload are provided as candidates. Four LLM clock-gating agents, each given a guideline for a different clock-gating technique, then sequentially refine the code. A code update is accepted only if it is functionally equivalent and reduces power without violating area and timing constraints. 

Experimental results with five RTL designs have demonstrated an average FGE improvement of 12.1\%p and an average power reduction of 23.6\% (up to 59.3\%) after CTS under the selected workloads, outperforming a commercial tool and an LLM baseline (4.1\% and 8.6\%). The results also show that the selected workload determines which registers are gated, and that a single agent given all guidelines neglects some techniques, which dedicated agents avoid.

%Several limitations remain. Although distinct workloads are generated, how well they represent actual usage scenario remains uncertain. Assigning each clock-gating technique to a separate agent improves refinement, but manually defining an agent and its guidance for each technique makes it difficult to extend the approach to other low-power techniques. Furthermore, equivalence checking and PPA evaluation incur substantial runtime overhead. Future work will address these limitations to improve the applicability and efficiency of LLM-assisted RTL refinement.

\bibliographystyle{ACM-Reference-Format}
\bibliography{references}

\end{document}